# Ultra-high critical current densities of superconducting YBa$_2$Cu$_3$O$_{7-\delta}$ thin films in the overdoped state

A. Stangl[1,*], A. Palau[1], G. Deutscher[2], X. Obradors[1], T. Puig[1,*]

[1]Institut de Ciència de Materials de Barcelona (ICMAB-CSIC) Campus de Bellaterra, 08193 Bellaterra, Barcelona, Spain

[2]Department of Physics and Astronomy, Tel Aviv University, 69978 Tel Aviv, Israel

* corresponding authors: teresa.puig@icmab.es, alexander.stangl@grenoble-inp.fr

## Abstract

Doping is one of the most relevant paths to tune the functionality of cuprates, it determines carrier density and the overall physical properties of these impressive superconducting materials. We present an oxygen doping study of YBa$_2$Cu$_3$O$_{7-\delta}$ (YBCO) thin films from underdoped to overdoped state, correlating the measured charge carrier density, $n_\mathrm{H}$, the hole doping, $p$, and the critical current density, $J_c$. Our results show a continuous increase of $J_c$ with charge carrier density, reaching 90 MA/cm² at 5 K for p-doping at the Quantum Critical Point (QCP), linked to an increase of the superconducting condensation energy. The ultra-high $J_c$ achived corresponds to a third of the depairing current, i.e. a value 60 % higher than ever reported in YBCO films. The overdoped regime is characterized by a sudden increase of $n_\mathrm{H}$, associated to the reconstruction of the Fermi-surface at the QCP. Overdoping YBCO opens a promising route to extend the current carrying capabilities of REBCO coated conductors for applications.

## Introduction

The advent and commercialisation of high temperature superconducting (HTS) cuprates can have a major impact on our modern urban societies and in the advancement of a low-carbon energy society, inevitable in the fight of climate change[1]. Important improvements have been made within the last 30 years since the discovery of high temperature superconductivity to tackle the most critical materials issues. The successful development of superconducting Coated Conductors (CC) is based on the uniform textured deposition of thick, homogeneous nanoengineered structures in kilometres length (a desired correlation over twelve orders of magnitude!). CC materials have opened up many new opportunities in the field of superconducting applications[2] and offer the capability to boost moonshot projects such as fusion reactors beyond ITER, dissipation-free energy transmission in superconducting grids or highly efficient engines for electrical aviation. However, current carrying capacities are still far from intrinsic limitations[3,4].

While on one hand dissipation free current transport is intrinsically limited by the depairing current density, $J_\mathrm{d}$, beyond which Cooper pair condensation is not energetically favourable anymore, in real systems the dissipation free, critical current density, $J_c$ is smaller due to the motion of magnetic vortices. One major goal in the development of superconducting tapes is to merge $J_c$ with the theoretical limit, $J_\mathrm{d}$. The main approach to increase $J_c$ nowadays is the introduction of nanoscaled defects (artificial pinning centres) in the YBCO matrix to immobilize magnetic vortices[5–10]. The total pinning force, $F_\mathrm{p} = J_c \times B$, is enhanced by an increased number of elementary pinning sites of optimised size and distribution [11]. However, the normal and superconducting (SC) state properties of cuprate materials are strongly governed by hole doping of the superconducting $CuO_2$-planes. Likewise, the elementary pinning strength itself varies with the condensation energy, $E_c$, per coherence volume which was predicted to peak at the critical doping $p^*$=0.19 where the pseudogap closes[12]. In [13] it was reported that the mean field value of the heat capacity jump at $T_\mathrm{c}$, and therefore the condensation energy, increases up to the maximum oxygen stoichiometry $O_7$, beyond optimal doping, in bulk YBCO. More recently, a strong increase of $E_c$, between the optimal doping ($p$=0.16) and $p*$ was found by measurements of the critical fields[14,15]. These results emphasize the possibility of enhancing $J_c$ by overdoping YBCO. However, although some attempts exist to obtain overdoped YBCO films[16], overdoping of YBCO films has been quite difficult to achieve up to now.

In this work, we have reached the overdoped state by oxygen post-processing of YBCO thin films with a thickness of 200 and 250 nm, grown from pulsed laser deposition (PLD) and chemical solution deposition (CSD), respectively. We show the influence of oxygen doping on the charge carrier concentration determined by Hall effect measurements all the way to the overdoped state. We find that $J_c$ is proportional to the charge carrier density, far into the overdoped state where a large Fermi surface of well-defined quasiparticles exists[17]. We demonstrate that overdoped YBCO films achieve record $J_c$ values of 90 MA/cm² at 5 K self-field, reaching a third of the depairing current density. Our results are in line with recent proposals of the controversial extrapolation of the pseudogap line, its relationship with the Quantum Critical Point and a Fermi Surface reconstruction at the critical point[18–20], which are crucial issues to understand the nature of high-$T_c$ superconducting cuprates, particularly the overdoped state. We foresee the hybridization of overdoping and nanoengineering of HTS-CC as an emerging opportunity to significantly improve CC performances for high magnetic field applications.

## Results

**Normal state electrical properties and doping state of YBCO thin films**

Normal state electrical properties of cuprate superconductors strongly vary with doping. Underdoped YBCO thin films were obtained by lowering the oxygen partial pressure below 1 bar. Overdoped YBCO films were achieved by post-growth oxygen heat treatments at low temperatures, enabled by catalytically activated surface oxygen exchange, using a thin Ag surface decoration layer (SI-Figure 1). The catalytic effect occurs via a job sharing mechanism where Ag dissociates the molecular oxygen and YBCO incorporates the oxygen ion into the bulk[21]. Electrical resistance and Hall measurements as a function of temperature have been employed in addition to room temperature *c*-parameter analysis to determine the doping state of the studied films. Representative examples of the temperature dependence of the Hall resistance $R_\mathrm{H}(T)$ and the electrical resistivity $\rho(T,H)$ at different magnetic fields are given in Figure 1 for our films. We have inferred the charge carrier density $n_\mathrm{H}(T) = \frac{1}{eR_\mathrm{H}(T)}$ by Hall measurements at constant magnetic field (3 T) and sweeping temperature (SI-Figure 2). We observe a strong temperature dependence of $R_\mathrm{H}$, respectively $n_\mathrm{H}$, as previously reported for HTS cuprates[22,23], with a maximum of $R_\mathrm{H}$ around 100K as shown in Figure 1(a). A similar drop of $R_\mathrm{H}(T)$ below 100 K in underdoped YBCO was recently explained by a reconstruction of the Fermi surface in the normal state due to charge-density-wave order[24,25]. However, in the present case the downturn in $R_\mathrm{H}$, which is also observed in optimally and overdoped YBCO, is caused by the onset of superconductivity, as the applied magnetic field ($\mu H_\mathrm{a} = 3$ T) is too weak to suppress the occurring phase transition to the superconducting state. In this work, we use the charge carrier density extracted at 100 K to study the correlation with superconducting properties, such as the critical current density.

The resistivity, $\rho(T,H)$ in Figure 1(b) reveals a broadening of the superconducting transition width with increasing magnetic field. From $\rho(T,H)$ measurements we have inferred the coherence length, $\xi(0)$, and a characteristic magnetic field, $H_0$, as discussed below. The deviation of $\rho(T)$ from a linear temperature dependence at low *T* is shown in the inset of Figure 1(b) which correlates with under- and overdoping, respectively, consistent with previous reports[26].

Figure 2 displays the critical temperature as a function of the Hall number, $n_\mathrm{H}(100\,\mathrm{K})$, measured at 100 K. $T_\mathrm{c}$ rapidly increases at low values of $n_\mathrm{H}$, as expected from the parabolic dependence of $T_c$ on doping generally found in cuprate superconductors[27] (assuming direct proportionality between the charge carrier density and doping in the underdoped region). Above the optimal doping of about $n_\text{H}(100\,\text{K}) = 3 \cdot 10^{21}/\text{cm}^3$, $T_c$ decreases, but following a much weaker dependence, as shown in the inset. This deviation from a parabolic dependence of $T_c$ was reported previously[26] and can be explained by an analysis of the doping dependence of the charge carrier density, as performed in the following.

In the underdoped regime, we calculated $p$ from measurements of $T_c$ using the parabolic dependence $1 - \frac{T_c}{T_{c,\text{max}}} = 82.6(p - 0.16)^2$, with $T_{c,\text{max}} = 92$ K, generally found for cuprates. However, this method was found insufficient for optimally and overdoped thin films. Hence, for these higher doped films we determined the doping number from measurements of the $c$-parameter by HR-XRD, as described in the methods (see also SI-Figure 3).

In Figure 3 we plot the evolution of the charge carrier density, $n$, at $T$=100K with doping, $p$. Here we use the charge carrier density per Cu in the CuO$_2$-planes, $n = \frac{n_\text{H} V}{2}$, with the volume of the unit cell $V$ (the factor ½ is owing to the fact that YBCO has two CuO$_2$-planes per unit cell). For $p$<0.16, we find $n = p$ in a broad doping range (underdoped regime). Above optimal doping ($p$>0.16), $n$ sharply increases. A similar result was reported previously, where $n_\text{H}$ was measured at low temperatures using very high magnetic fields suppressing the superconducting state (e.g. at 50 K and up to 88 T)[24]. This can be understood by a Fermi surface reconstruction (FSR) in proximity to the pseudogap (PG) critical point at $p$*, resulting in a non-unique relation between the charge carrier density, $n$, and doping, $p$, over the full range of the cuprate phase diagram[20]. In the far overdoped regime, cuprate superconductors have been recently shown to exhibit a large cylindrical Fermi Surface (FS) with $n(T \to 0) = 1 + p$ in the zero temperature limit, as described in [24,25,28,29]. Below the critical doping, the volume of the FS reduces by one hole per Cu in the CuO$_2$ plane to $n(T \to 0) = p$, e.g. due to the introduction of an antinodal gap opening which seems to arise from short range antiferromagnetic correlations in the system[18]. This transition is expected to occur within a narrow doping range between the optimal doping $p = 0.16$ and the closing of the PG at the critical doping $p^* = 0.19$, as observed in our study. The PG critical point at $p^*$ has been shown to have all the features of a quantum critical point at $T = 0$ [30].

It is remarkable that the charge carrier density measured at 100 K, above the onset of superconductivity, preserves the expected behaviour for $n(T \to 0)$ of a single band metal with a Fermi-surface containing small hole-like pockets on the underdoped site and a sharp transition towards a large Fermi-volume, due to a reconstruction of the Fermi surface, in overdoped YBCO. The latter is also the reason for the observed deviation from a parabolic doping dependence of $T_c$ on $n_\mathrm{H}$, shown in Figure 2.

**Superconducting properties of YBCO films: Correlation with doping and condensation energy**

The main result of this work is shown in Figure 4, where we plot the critical current density, $J_\mathrm{c}$, as a function of $n_\mathrm{H}(100\,\mathrm{K})$ in (a) at self-field and (b) at an applied magnetic field of 7 T, determined by magnetisation measurements at 5 K (SI-Figure 4). For both cases, we find a strong quasi-linear increase of $J_c$ with $n_\mathrm{H}$, extending beyond optimal doping, far into the overdoped regime up to the equivalent value of *p**=0.19 (corresponding $n_\mathrm{H}$ values for optimal and critical doping are indicated with vertical dashed lines). Notice the ultrahigh values of $J_c$ at self-field and 5 K, $J_c(5\,\mathrm{K})$ achieved beyond optimal doping up to *p**.

An evaluation of the depairing current within the Ginzburg-Landau theory[31] is done in the following in order to compare the obtained experimental values with this theoretical limit, though strictly valid only near $T_c$:

$$J_\mathrm{d}^\mathrm{GL}(T) = \frac{\phi_0}{3^{3/2}\pi\mu_0\lambda^2(T)\xi(T)} \qquad \text{(Eq. 1)}$$

with the temperature and material dependent magnetic penetration depth $\lambda(T) = \lambda(0)/\sqrt{1-t^4}$ and coherence length $\xi(T) = \xi(0)\sqrt{(1+t^2)/(1-t^2)}$, the flux quantum $\phi_0$ and the reduced temperature $t = T/T_c$. Using $\xi(0) = 1.6\,\mathrm{nm}$ and $\lambda(0) = 140\,\mathrm{nm}$ for YBCO, we can estimate $J_d^{GL}(5\,K) \approx 300\,MA/cm^2$. The maximum $J_c(5\,K)$ obtained within this study is 89.4 MA/cm², and thus close (factor 3) to the fundamental limit. To the best of the authors knowledge, this is the highest ever-reported $J_c(5\,K)$ value for thin film superconductors, exceeding by more than 60 % previous record values of REBCO at zero field[32,33]. We suggest, that this linear increase of $J_c$ in YBCO films with the charge carrier density up to $p^*$ is a consequence of the linear increase of the condensation energy with charge carrier density, as shown in the following.

Vortex matter in HTS cuprates has demonstrated to be extremely rich with new vortex phases that were not expected from the low temperature superconductivity knowledge[34–36], specially related to the high thermal energy and flux creep phenomena[37] that these cuprates experience. In this context, the pinning energy (which is proportional to the condensation energy, $E_c$ [32]) can be related to an effective activation energy determined from magnetoresistance measurements, which can be written generally as $U(H,T) = \left(1 - \frac{T}{T_c}\right)^m \left(\frac{H_0}{H}\right)^\beta$ [38,39], where $\beta$ is a constant close to unity and $m$ a material dependent parameter. The characteristic magnetic field $H_0$ is proportional to the pinning energy, and thus closely related to the condensation energy [36,40,41]. We have obtained $H_0$ by analysing the vortex glass transition line, given by the empirical formula $H_G = H_0 \left[\frac{1-t(H)}{t(H)}\right]^{1/\beta}$ [42–44], with $t(H) = T_G(H)/T_c$, where $T_c$ is the critical temperature at zero field and $T_G(H)$ the field dependent vortex-glass transition temperature. We find that $H_0$ depends on the Hall number, as shown in Figure 5(a). As $n_H$ increases from 2 to $9 \cdot 10^{21}$/cm³, $H_0$ doubles from 35 to > 70 T. Thus, suggesting an increase of the condensation energy, $E_c$, of the order of a factor two within this doping range. This is in agreement with the trend observed from the analysis of the condensation energy, $E_c$, determined from measurements of the specific heat jump in Bi2212[45] and YBCO[46], and the upper and lower critical fields, $H_{c1}$ and $H_{c2}$ in YBCO[14], which can be correlated, according to BCS theory, to the zero temperature condensation energy per unit volume, $E_c \propto \Delta C$ and $E_c \propto H_c^2$. In these two cases, $E_c$ also approximately doubles from optimal to the critical doping[14,41]. In the inset of Figure 5(a) we report a linear dependence of $J_c(5\,\text{K})$ on $H_0$. The fact that these two quantities, obtained by two independent techniques, correlate perfectly, demonstrates the importance of the condensation energy, $E_c$, as the underlying quantity governing both parameters, and gives consistency for $H_0 \propto E_c$ in the above analysis.

To verify the plausibility of our results, we analyse the expected relationship between the self-field critical current density, $J_c$, the condensation energy, $E_c$ and the charge carrier density, $n_\text{H}$. Therefore, we consider the condensation energy per pair $u(T)$, which is equal to the critical kinetic energy at zero temperature:

$$u(0) = \tfrac{1}{2} m v_c^2, \qquad \text{(Eq. 2)}$$

with the critical velocity of a pair $v_c$. The condensation energy per pair is equal to the condensation energy per unit volume divided by the superconducting pair density $n_s$, hence $u(0) = E_c(0)/n_s$. Using the critical velocity, we can generally write the critical current density as

$$J_c = e(2n_s)v_c. \qquad \text{(Eq. 3)}$$

Combining Equation (2) and (3) we obtain

$$J_c^2 \propto n_s E_c. \qquad \text{(Eq. 4)}$$

We expect the square of the critical current density to vary as the superconducting pair density multiplied by the condensation energy. We further assume that $n_s$ varies with the measured Hall carrier density $n_\text{H} \propto 2n_s$. Going from optimal doping to the highest achieved over-doping, $n_H$ increases by a factor of 3, while $H_0$ increases by a factor of 2. From the derived expression, the critical current density is expected to increase over that range by a factor of 2.5, which is consistent with our results reported in Figure 4(a). In fact, using the experimental result $H_0 \propto E_c \propto n_\text{H}$ shown in Figure 5(a), Eq. 4 transforms to Eq. 5, which is experimentally verified in Figure 4, thus completing the consistency of the results in the doping range analysed,

$$J_c \propto n_s. \quad \text{(Eq. 5)}$$

Thus, we believe that the increase of $J_c$ with the charge carrier density is able to confirm an increase of the condensation energy in YBCO films with ultrahigh critical currents up to $p^*$. We have therefore been able to confirm the increase of $E_c$ from a magnitude in the superconducting state ($J_c$), being in agreement with previous results determined from specific heat[41] and critical fields[14] measurements. Some of these works have suggested the existence of a peak in the condensation energy $E_c$, and thus in the density of states, $N_F$, at $p^*$ when the PG closes and cuprates enter the strange metal state before reaching the Fermi liquid behaviour[12,14,20,41]. This is one of the features of a QCP irrespective of the details of the microscopic model used to describe the PG formation[20,40,47,48]. Unfortunately, this has been measured up to now only in Ca-doped YBCO and $La_{1.8-x}Eu_{0.2}Sr_xCuO_4$[20]. It would be very interesting to be able to overdope these YBCO films beyond the present $p^*$ value to confirm also the peak in $E_c$.

There is still some controversy in the field about the origin of the onset of self-field dissipation[49,50], e.g. whether it is a Silsbee or pinning dominated mechanism. However, as within this work, $J_c$ was obtained via the measurement of the remanent magnetisation, and specially the linear relationship demonstrated in Eq. 5, $J_c \propto n_s$, also applies for 7 T (Figure 4(b)), we claim that in this case pinning plays a crucial role. Hence, variations of the coherence length might contribute to the increase of the critical current, as the pinning energy, $U_p(0) = E_c(0)\xi(0)^3$, scales with $\xi(0)^3$ [41] and the coherence length itself exhibits a dependence on doping[51–53]. To be able to exclude that a major contribution arises from changes in $\xi(0)$, we have analysed $\xi(0)$ for several samples spanning a broad doping range, as shown in Figure 5(b). $\xi(0)$ is extracted from electrical measurements of the flux flow resistivity up to 9 T as described in [54–56]. We observe only a small variation of $\xi(0)$, while the Hall number changes by more than a factor 4. Thus, the $J_c$ enhancement due to this modified coherence length would account for an upper limit of 25 %, much less than the observed increase of >300 % within the doping range from 2 to $8 \cdot 10^{21}$/cm³. Thus, we conclude that the enhancement of $J_c$ is mainly governed by the modified condensation energy due to the increase of oxygen doping.

**Consequences in pinning behaviour and comparison with nanoengineered films**

YBCO films are very sensitive to atomic and nanoscale defects due to the small coherence length in high temperature superconductors, which in addition leads to strong thermal fluctuations, especially at high temperatures, which strongly hinder vortex pinning. The consequence is that very different behaviour is observed for different pinning centres at different temperatures. While columnar defects are more advantageous at high temperatures, weak uncorrelated pinning sites strongly contribute to the overall pinning force at reduced temperatures[32,57]. The effect of random point defects is observed to be more significant at low temperatures[7]. $J_c(77\,\text{K})$ vs $J_c(5\,\text{K})$ is shown in Figure 6 for films grown by PLD and CSD. In both types of samples, we observe a strong correlation between $J_c$ at the two different temperatures up to about $J_c(5\,\text{K}) = 50\,\text{MA/cm}^2$. However, CSD and PLD films follow different trends, which may be attributed to different pinning efficiencies associated to two different microstructures. CSD films typically preserve a much stronger distorted matrix, giving raise to strong pinning in strained regions, being more efficient at higher temperatures[9]. Above $J_c(5\,\text{K}) = 50\,\text{MA/cm}^2$, PLD films show a deviation from the initial linear relation with $J_c(77\,\text{K})$, resulting in a much weaker dependence ($J_c(77\text{K})$ increases from 3 to 4 MA/cm², while $J_c(5\,\text{K})$ almost doubles from 50 to 90 MA/cm²). We propose that this saturation is caused by the reduced efficiency of weak pinning sites at high temperatures. Notice, in Figure 6 upper panel, the consistency of $T_c$ with self-field $J_c(5\,\text{K})$, further demonstrating the intrinsic relationship of these two magnitudes with the charge carrier density.

The strong increase of $J_c$ with doping motivates for a comparison of the in-field $J_c(H)$ properties with reported values from other optimized approaches, including the more recent nanoengineering of the microstructure by the embedding of nanoparticles and nanorods in HTS films and coated conductors. $J_c(H \parallel c)$ at low temperatures is shown in Figure 7 for various different samples (see Figure caption for details on sample composition). The most striking feature is the strong enhancement of $J_c$ at self-field ($\mu_0 H = 0$) in the overdoped film, compared to conventional approaches by nanodefecting the matrix, as discussed above. Up to $\mu_0 H = 2\,\text{T}$, overdoping enables the highest ever reported $J_c(H)$, with an improvement of up to 60 % at self-field. Weak pinning, the expected main contribution in pristine YBCO films, is known to rapidly decrease with small magnetic fields[7], explaining the rapid decrease of $J_c(H)$ below 1 T for the overdoped PLD sample. However, the extraordinary high self-field $J_c$ asserts high $J_c(H)$ at even intermediate fields, comparable to the best performing nanocomposites, whose pinning is governed by 1D nanorods and strained regions around 3D nanoparticles. We also show a representative example of a pristine optimally doped YBCO film, as reported in [58], highlighting the potential of overdoping YBCO. We notice that in this study, $J_c$ was obtained by magnetisation measurements, which typically results in slightly smaller values compared to electrical transport measurements due to flux creep effects, making our results even more remarkable. Therefore, we encourage future work achieving the overdoped state in nanoengineered films to reach ultrahigh critical current densities also at high magnetic fields.

## Discussions

We have fabricated overdoped YBCO thin films by means of different post-processing oxygen heat treatments being able to tune the doping state and reach the critical doping value *p**=0.19, close to the Quantum Critical Point. The overdoped state is confirmed by a small decrease of $T_c$ and a transition of the normal state charge carrier density from $n = p$ to $n = 1 + p$, where *p* is the doping state of the $Cu_2O$ planes, in agreement with the proposed reconstruction of the Fermi surface above optimal doping (*p*=0.16). However, this result is remarkable, as $n$ was obtained above the onset of superconductivity at 100 K, preserving the expected behaviour from the limit $T \to 0$. We suggest that low temperature measurements at high fields, necessary to suppress superconductivity, would be highly interesting to reveal the further temperature evolution of $n_H(T)$ in these thin films.

The evaluated overdoped regime is characterised by an increase of the condensation energy, leading to extraordinary self-field $J_c$ values at low temperatures, up to one third of the depairing current, reaching 90 MA/cm² at 5 K. We demonstrate a general linear increase of $J_c$ with the charge carrier density, $n_H$, at self-field and high magnetic fields of 7 T up to the critical point p*. The distinct behaviour of $J_c$ observed at 5 K and 77 K, suggests that we have been able to modify the weak pinning individual strength through the modification of the condensation energy by doping. p-doping strategies with oxygen post-processing treatments are expected to be scalable and uniform in long length, therefore, we envisage a viable hybridisation of overdoping and nanoengineering of YBCO films, which offers powerful prospects to further push prevailing limitations of dissipation-free current transport in cuprate superconductors and Coated Conductors at high magnetic fields of interest for applications. We reinforce the interest to find ways to overdope YBCO films beyond the present critical value p* to confirm the existence of a peak in condensation energy, $E_c$, at higher doping levels and shine light on the consequences of crossing the Quantum Critical Point.

## Methods

**Film Fabrication**. The $Y_1Ba_2Cu_3O_{7-\delta}$ thin films are grown using chemical solution deposition (CSD) and pulsed laser deposition (PLD) on 5x5 mm² $LaAlO_3$ (100) and $SrTiO_3$ (100) single crystal substrates with thicknesses of 200 nm (PLD) and 250 nm (CSD), respectively. In case of CSD, the stoichiometric amount of precursor metal trifluoroacetate salts is dissolved in an alcoholic solution and deposited by spin-coating, followed by a pyrolysis (~300 °C) and growth (~800 °C) temperature treatment at $P_{O2}$=0.2 mbar[59,60]. PLD layers are deposited at 800 °C at $P_{O2}$=0.3 mbar with a pulse frequency of 5 Hz[61]. The PLD-targets were fabricated by Oxolutia SL (Spain) and consist of pressed and sintered, stoichiometric YBCO powder at 87 % density. After growth, a 100 nm thick surface decoration layer of patterned Ag is deposited on the surface by sputtering at RT (SI-Figure 1(a)). This Ag layer catalytically enhances oxygen exchange activity of YBCO during the following oxygenation process, as a Job sharing mechanism facilitates the dissociation and incorporation of $O_2$ (SI-Figure 1(b))[21,62]. Additionally, the silver coating provides good electrical contact for electrical measurements. Dewetting of the Ag layer into small islands with diameters of about 0.1 µm is observed at around 300 °C. However, this effect is not detrimental to its catalytic activity nor electrode functionality. Hole doping of YBCO is achieved by oxygen incorporation during the post growth oxygenation at 1 bar with an oxygen flux density of 0.16 l min⁻¹cm⁻² at different intermediate temperatures (280-550 °C) with dwell times between 30 and 240 min.

**Structural characterisation.** Layers obtained by either growth technique are epitaxial textured, twinned and highly c-axis oriented with no trace of secondary phases, as determined by X-ray diffraction measurements (Bruker D8 Discover), as shown in SI-Figure 3a. In the presented films with thicknesses above 200 nm, no macroscopic strain due to lattice mismatch with the used substrates ($LaAlO_3$ and $SrTiO_3$) was observed.

The *c*-lattice parameter is obtained by HR-XRD measurements using the Nelson-Riley method, which allows the determination of the lattice parameter with very high precision, as aberration errors are minimised at very high angles (see SI-Figure 3(b)). For optimally and overdoped films, we determined the doping number via the *c*-parameter, using the empirical equation $p = c_1 y + c_2 y^6 + p_0$ with $y = 1 - c/c_0$ [63,64]. The prefactors $c_i$ depend on the sample type and growth process. In this work we have used values reported in [63] ($c_0 = 11.8447$, $c_1 = 11.491$, $c_2 = 5.17 \cdot 10^9$). A small systematic constant offset was corrected by introducing the additional parameter $p_0 = -0.02$.

**Magnetic and electrical analysis.** SQUID magnetometry (Quantum Design) was used to determine critical current densities via the width of the magnetisation loop, as shown in SI-Figure 4(a), using the Bean critical state model for thin discs.

In-depth electrical analysis was performed using a Physical Property Measurement System (Quantum Design) over a broad temperature range. Contacts for electrical measurements were glued with silver paint on top of 400 µm squared Ag electrodes sputtered at the corners of the films. Electrical measurements were performed in Van der Pauw and Hall configuration in fields up to 9 T ($H \parallel c$), averaged over two permutations of the electrical contacts and positive and negative excitation current in DC mode. The studied YBCO films are highly twinned, therefore the influence of metallic CuO chains short circuiting the Hall voltage along the *b*-direction is neglected in the calculation of the Hall coefficient $R_\mathrm{H}$. The critical temperature at zero field, $T_c$, and the field dependent vortex-glass transition temperature, $T_\mathrm{G}(H)$, are determined from $\rho(T,H)$ measurements to the point where the electrical resistance in response to a small excitation current vanishes. The vortex-glass transition line $H_G = H_0 \left[\frac{1-t(H)}{t(H)}\right]^{1/\beta}$ is used to obtain the characteristic magnetic field, $H_0$, by linear fitting as shown in SI-Figure 5(a). From the same measurements, we were able to determine the upper critical field, $H_{c2}(T)$, and in the zero temperature limit, $H_{c2}(0)$, using the classical Werthamer-Helfand-Hohenberg relation, which in turn defines the coherence length, $\xi(0)$. The determination of $H_{c2}(T)$ is shown in SI-Figure 5(b).

## Data availability

The data that support the findings of this study are available from the corresponding authors on reasonable request.

## Acknowledgements

The authors acknowledge financial support from Spanish Ministry of Economy and Competitiveness through the "Severo Ochoa" Programme for Centres of Excellence in R&D (Grant No. SEV-2015-0496), ULTRASUPERTAPE (ERC-2014-ADG-669504), EUROTAPES project (FP7-NMP-Large-2011-280432CONSOLIDER Excellence Network (Grant No. MAT2015-68994-REDC), COACHSUPENERGY project (Grant No. MAT2014-56063-C2-1- R and SuMaTe RTI2018-095853-B-C21, cofinanced by the European Regional Development Fund), and from the Catalan Government with Grant No. 2014-SGR-753 and 2017-SGR-1519. Authors also thank the network collaboration of EU COST action NANOCOHYBRI CA16218.We also acknowledge the Scientific

Services at ICMAB. We thank Juri Banchewski for some of the last transport experiments. A.S. is grateful for illuminating discussions on the manuscript with Stephan Steinhauer (KTH, Sweden).

## Author Contributions

A.S., A. P. and T.P designed the experimental study. A.S. performed the experimental work and analysed the data. A.S. prepared the manuscript with contributions from co-authors. All authors contributed to the scientific discussion.

## Competing interest

The authors declare no competing interests

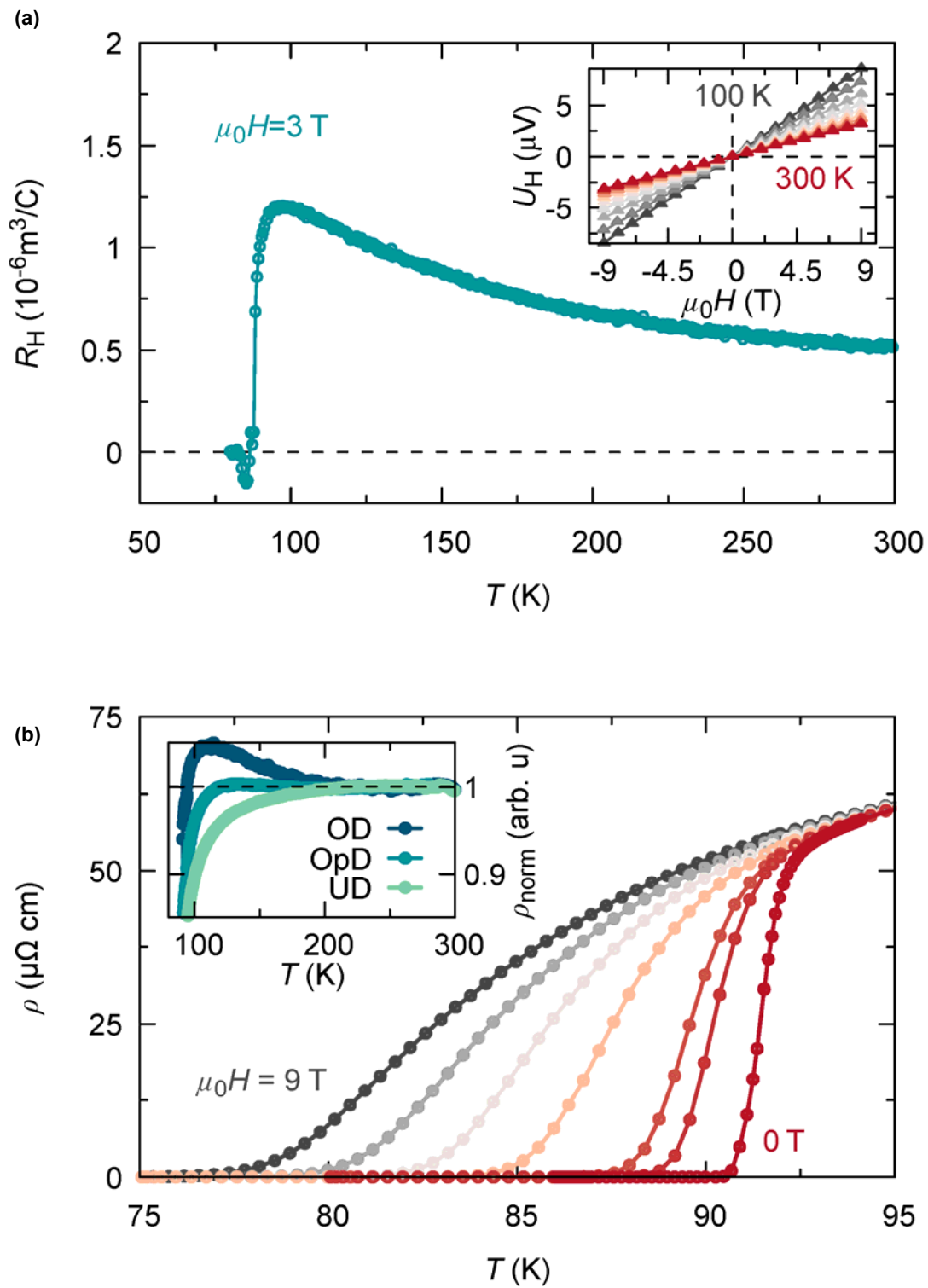

Figure 1: **Electrical analysis of YBCO thin films**: (a) Hall constant $R_H$ obtained at 3 T as a function of temperature. $R_H$ is perfectly linear within the analysed field range (-9 to 9 T) and vanishes in zero field condition at all $T$ as shown in the inset. Charge charrier

density is obtained via Hall effect measurements using $n_H(T) = \frac{1}{R_H(T)q}$. (b) In-plane resistivity, $\rho(T, H)$, as a function of temperature at different magnetic fields (0, 0.5, 1, 3, 5, 7, 9 T, $H \parallel c$) to evaluate field dependent superconducting transition temperature, $T_0(H)$. Inset shows normalized resistivity, $\rho_{\text{norm}} = (\rho(T) - \rho_0)/bT$, where $b$ is the linear slope at high temperatures ($T > 150$ K). Doping dependent deviation from unity is observed, as shown for an underdoped (UD, downwards deviation), optimally doped (OpD) and overdoped (OD, upwards bending) 200 nm thick YBCO film.

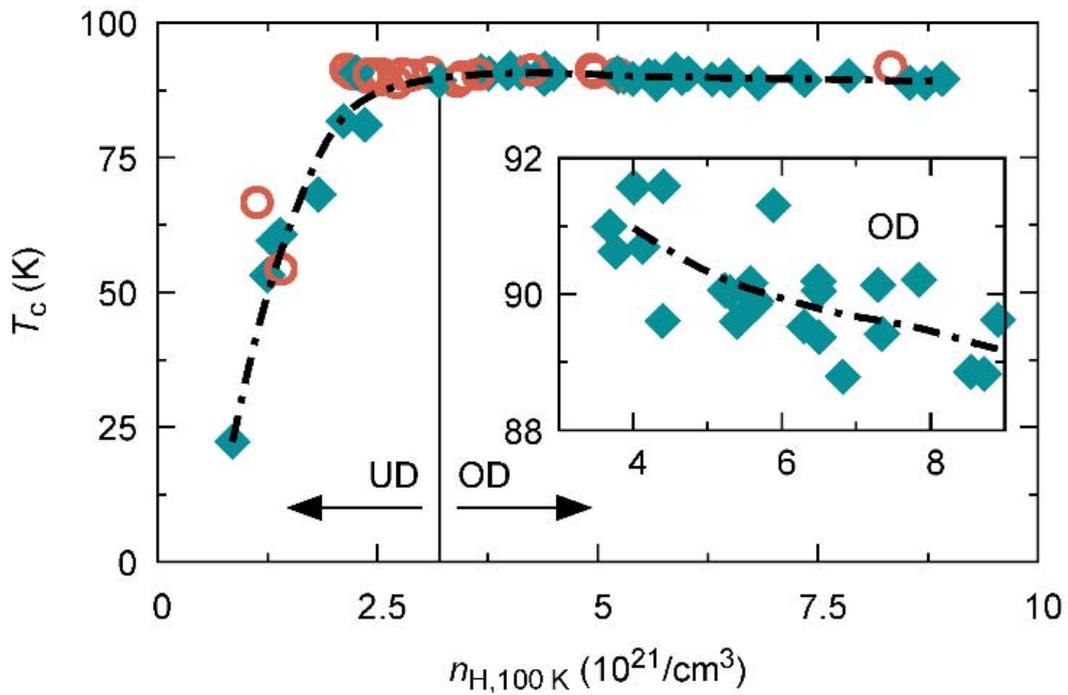

Figure 2: **Phase diagram of YBCO thin films:** Zero field critical temperature as a function of charge carrier density, $n_H$, obtained by Hall effect measurements at 100 K for thin YBCO films grown by PLD (200 nm thick, cyan diamonds) and CSD (250 nm thick, red circles). Vertical line marks optimally doping, while arrows indicate underdoped (UD) and overdoped (OD) regime. Inset magnifies $T_c$ in the overdoped regime, showing a weak but distinct decrease with increasing charge carrier density, from above 91 K to around 89 K.

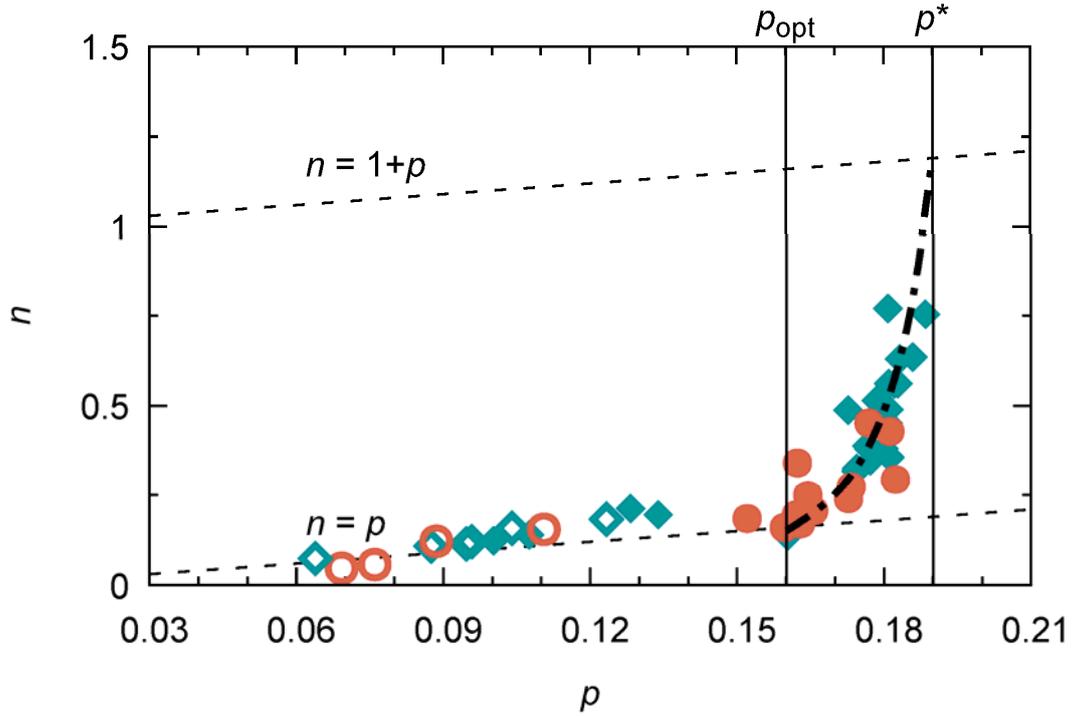

Figure 3: **Evolution of the charge carrier density with doping.** YBCO normal state charge carrier density per $CuO_2$-plane, $n$, is drawn as a function of doping $p$. The charge carrier density is given by $n = \frac{n_H V}{2}$ at 100 K. The doping $p$ is obtained for optimally and overdoped samples (full symbols) via HR-XRD measurements of the $c$-parameter and for underdoped films (open symbols) from the parabolic doping dependence of $T_c$. The vertical lines indicate optimally and critical doping. Below $p_{opt}$ we find $n = p$, corresponding to a Fermi-surface with small hole and/or electron pockets in the underdoped regime. For $p > p^*$ a large, cylindrical FS is expected in the metallic overdoped regime, with $n = 1 + p$. A transition between a small and a large Fermi-surface occurs above $p = 0.16$. This is in good agreement with previous reports, but remarkable, as within this work $n$ is obtained by Hall effect measurements using small fields above the onset of superconductivity at 100 K.

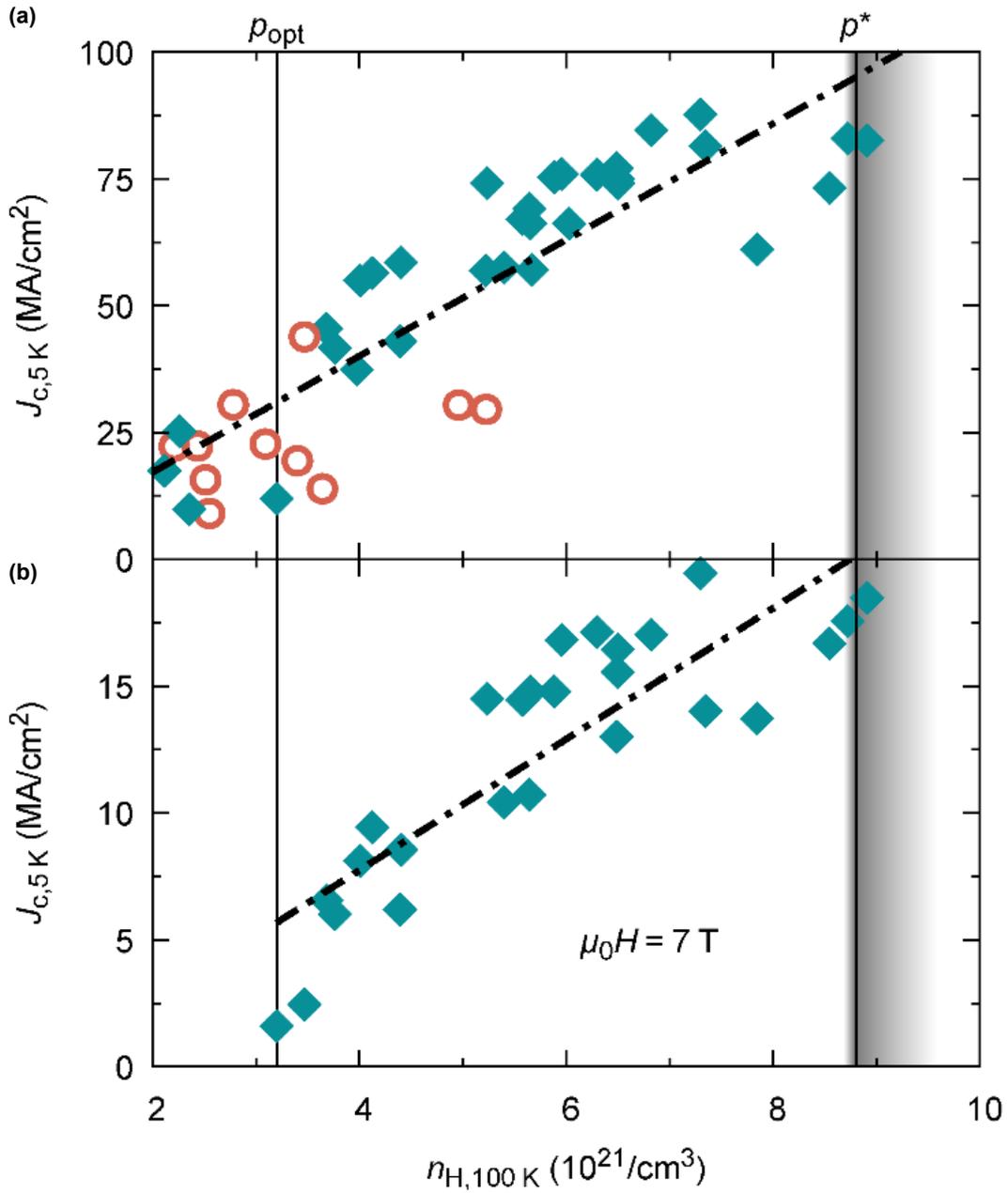

Figure 4: **Dependence of $J_c$ on charge carrier density:** Self-field critical current density, $J_c$, at 5 K versus charge carrier density $n_H(100\,\text{K})$ in (a) self-field and (b) an applied magnetic field of 7 T of YBCO thin films obtained by CSD (red circles, 250 nm) and PLD (cyan diamonds, 200 nm). The critical current density is determined by SQUID magnetisation measurements. $J_c$ is strongly enhanced by increasing the charge carrier density and approximately follows a linear relation far into the overdoped regime. Optimal and critical doping, $p_{opt}$ and $p^*$, are marked with vertical lines, while the shadowed area

around $p^*$ indicates the uncertainty of defining the critical doping in terms of $n_\mathrm{H}$ via Figure 3. Dashed lines in (a) and (b) are guides to the eye.

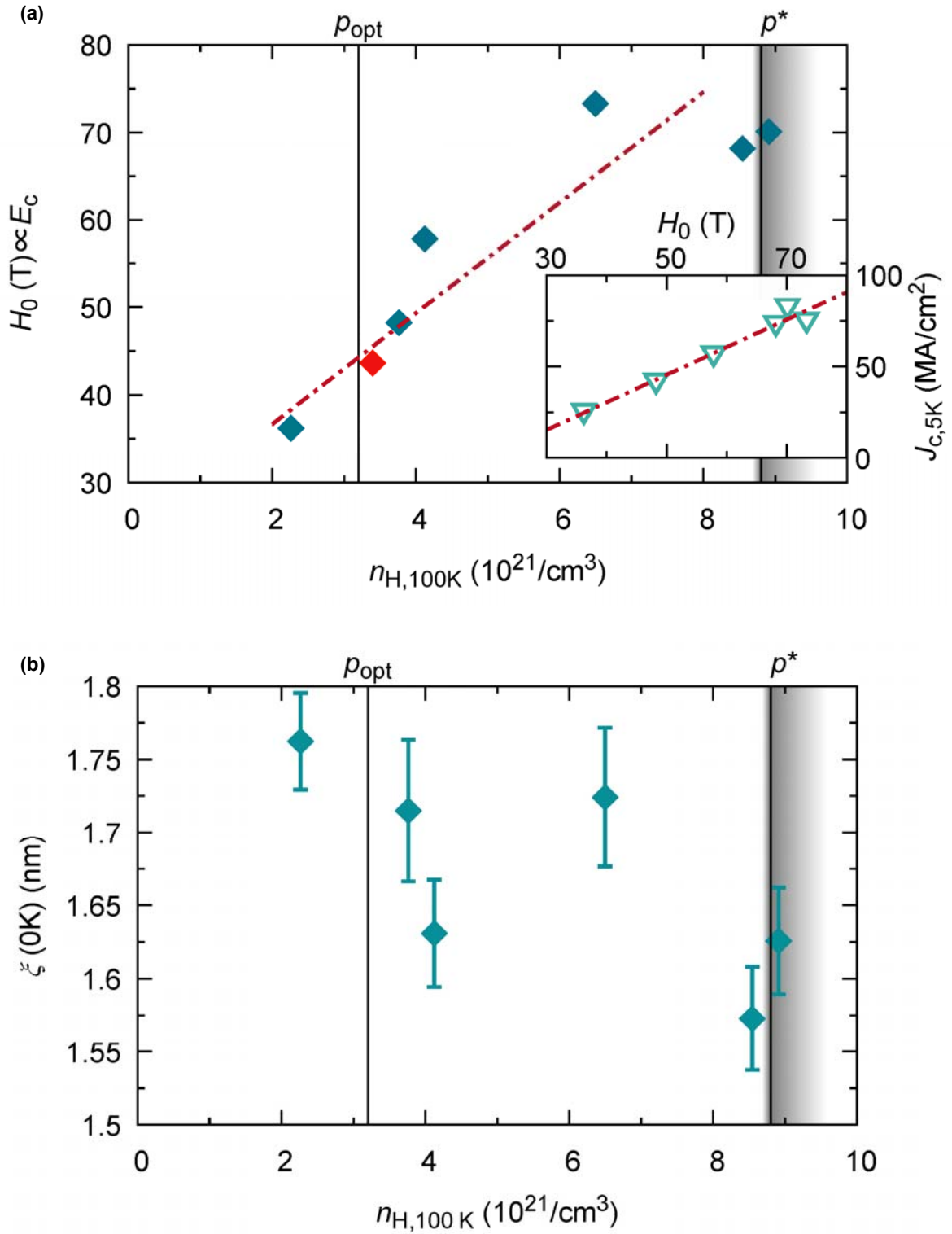

Figure 5: **Charge carrier density dependence of superconducting parameters:** (a) Characteristic magnetic field, $H_0$, versus charge carrier density at 100 K. $H_0$ is obtained

by electrical measurements of the vortex glass transition line. It provides a measure of the pinning energy and thus is closely linked to the superconducting condensation energy, $E_c$. The red diamond is reproduced from [40] (280 nm, PLD), falling on the same line as our results. Inset shows $J_c(5\,\text{K})$ versus $H_0$, revealing a linear relation between these independently measured quantities. This emphasizes the strong correlation of the critical current density and the condensation energy. Dashed lines are guides to the eye. (b) Superconducting coherence length $\xi(0)$ as a function of Hall number at 100 K. $\xi(0)$ is obtained by electrical measurements of $H_{c2}$, determined from flux flow resistivity analysis up to 9 T (SI-Figure 5(b)). A weak decrease with increasing charge carrier doping is observed, which alone cannot account for the strongly enhanced $J_c$ in the overdoped region.

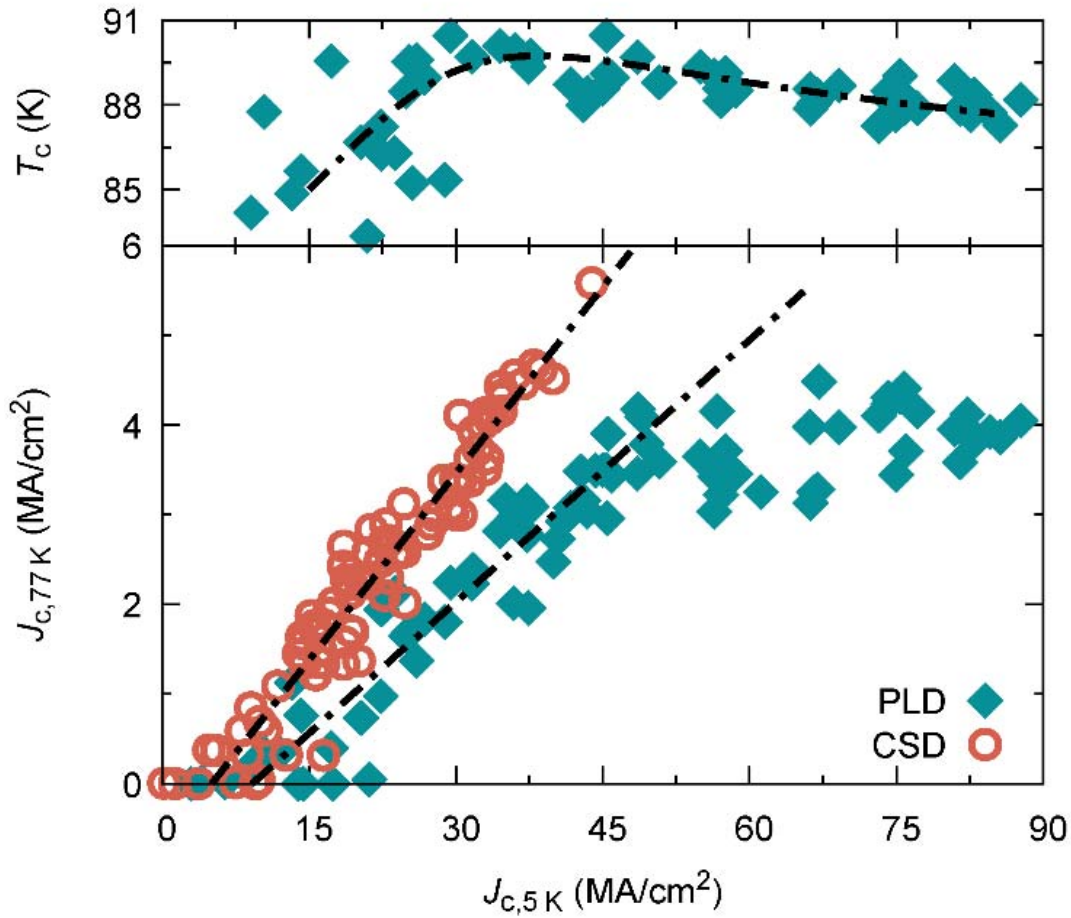

Figure 6: **Superconducting physical properties correlations for PLD and CSD films:** $T_c$ dependence with self-field $J_c(5\,\text{K})$ (upper panel) showing the intrinsic relation of both quantities with the charge carrier density. Self-field $J_c(77\,\text{K})$ as a function of self-field $J_c(5\,\text{K})$ for YBCO films obtained by PLD and CSD (lower panel). The different linear trends for $J_c(5\,\text{K})<50\,\text{MA/cm}^2$ reveal growth dependent pinning defect landscapes, with possibly higher strong-pinning contribution in CSD films. With increasing $J_c(5\,\text{K})$ in PLD films, $J_c(77\,\text{K})$ saturates, probably due to a reduced contribution of weak pinning at high temperatures and the proximity to the superconducting transition in the overdoped regime, as indicated in the upper panel by a decreased $T_c$ for high $J_c(5\,\text{K})$ films. All lines are guide to the eye.

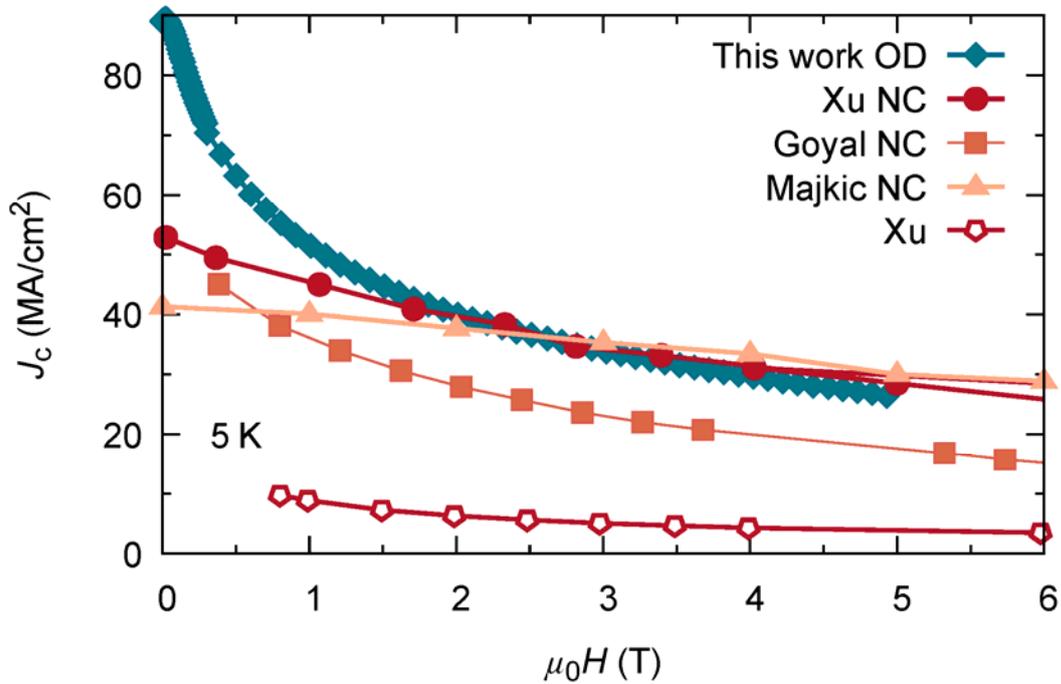

Figure 7: $J_c(H \parallel c)$ **at 5 K of some best performing YBCO films:** Field dependence of reported high critical current densities for several nanocomposite (NC) YBCO thin films in comparison with overdoped (OD) YBCO obtained in this work (PLD). Overdoped YBCO exhibits a remarkable self-field $J_c$, almost 60 % higher than previous record films, compensating the fast decrease of $J_c(H)$ at low fields as typical for pristine YBCO films. From literature reproduced results cover the currently best-practice strategies of nanoengineering YBCO of coated conductors: 15 % Zr doped (Gd,Y)BCO (Xu NC, MOCVD, at 4.2 K, reproduced under CC-BY)[33], nanoscale defected REBCO with 4 %

BZO (Goyal NC, PLD, at 5 K, reproduced under CC-BY)[65], REBCO with 15 % Zr addition (Majkic NC, MOCVD, at 4.2 K, raw data was kindly provided by the author)[66]. Additionally we show a pristine YBCO film (Xu, MOCDV, at 4.2 K, reproduced with permission from AIP publishing) as reported in [58].

# Supplementary Information

## Ultra-high critical current densities of superconducting YBa$_2$Cu$_3$O$_{7-\delta}$ thin films by oxygen overdoping

A. Stangl[1,*], A. Palau[1], G. Deutscher[2], X. Obradors[1], T. Puig[1,*]

[1]Institut de Ciència de Materials de Barcelona (ICMAB-CSIC) Campus de Bellaterra, 08193 Bellaterra, Barcelona, Spain

[2]Department of Physics and Astronomy, Tel Aviv University, 69978 Tel Aviv, Israel

* corresponding authors: teresa.puig@icmab.es, alexander.stangl@grenoble-inp.fr

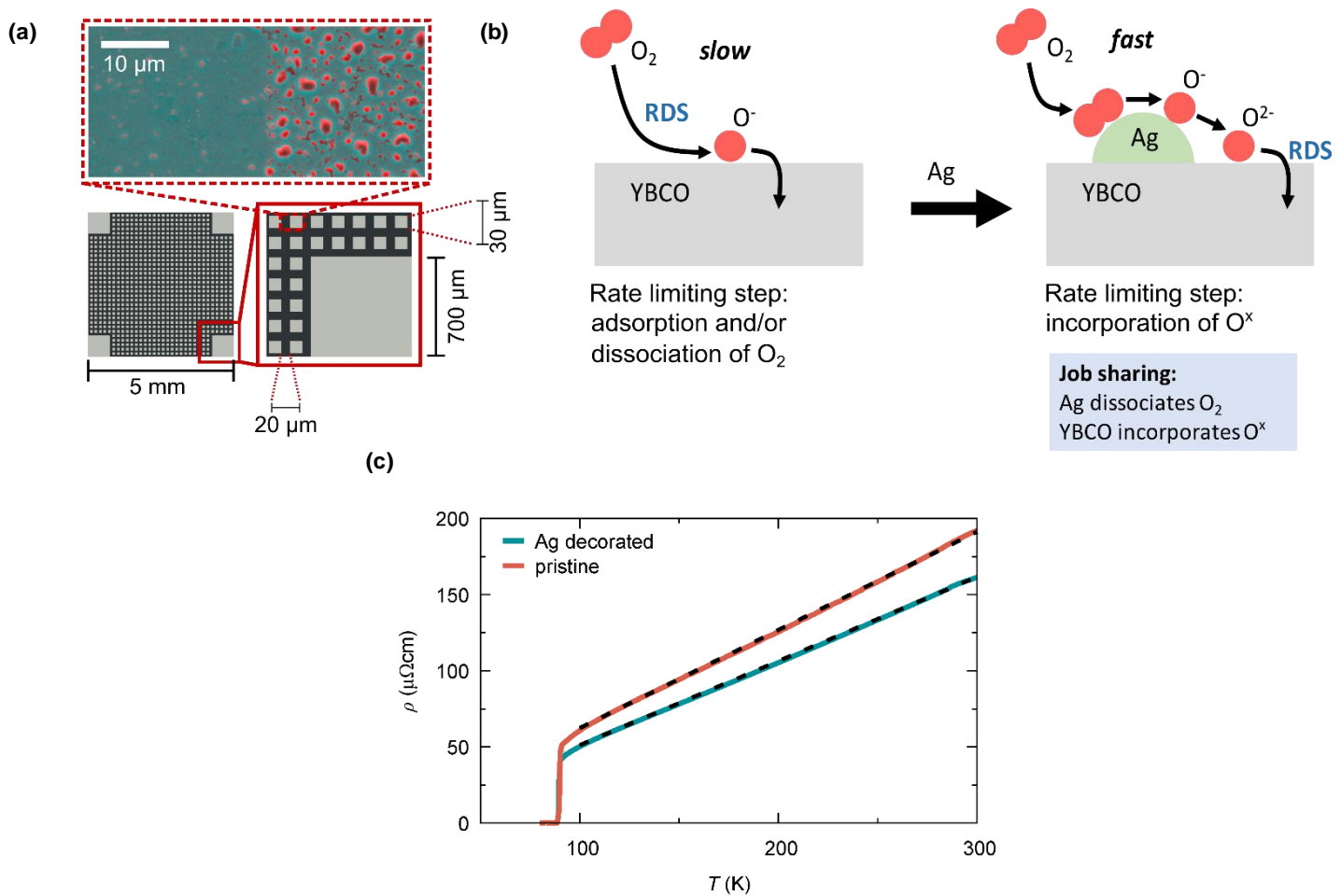

*SI-Figure 1:* **Role of Ag in post oxygen heat treatment:** (a) Schematic of silver surface decoration using a 100 nm thick mesh of 30x30 µm squares on top of the superconducting film surface and 700 µm big pads in the corners used as electrodes for electrical measurements (as-deposited Ag surface coverage: 40 %). False colour SEM image in combined SE (secondary electron) and BSE (back-scattered electron) mode in the top panel shows dewetting of continuous layer into Ag islands as observed after heat treatments above 300 °C, reducing the coverage to about 10 %. (b) Role of Ag during the oxygen incorporation process. In pristine YBCO thin films oxygen exchange is rate limited (rate determining step, RDS) by adsorption and/or dissociation of molecular oxygen onto the film surface, resulting in a sluggish overall oxygen reduction reaction (ORR). Catalytic activity of Ag strongly accelerates these processes and oxygen is efficiently incorporated into the YBCO bulk at the triple phase boundary around the Ag particles even at low temperatures, enabling low temperature oxygen post treatments[1]. (c) Electrical resistivity measurements of Ag decorated and pristine YBCO 200 nm thick films, showing that Ag surface decoration layer does not affect electrical measurements. The small difference in $\rho(T)$ is caused by different charge carrier densities, with the Ag decorated one having a $n_\mathrm{H}(100\,K)$ of $7.4 \times 10^{21}$ /cm³ in comparison to $5.7 \times 10^{21}$ /cm³ in the pristine film.

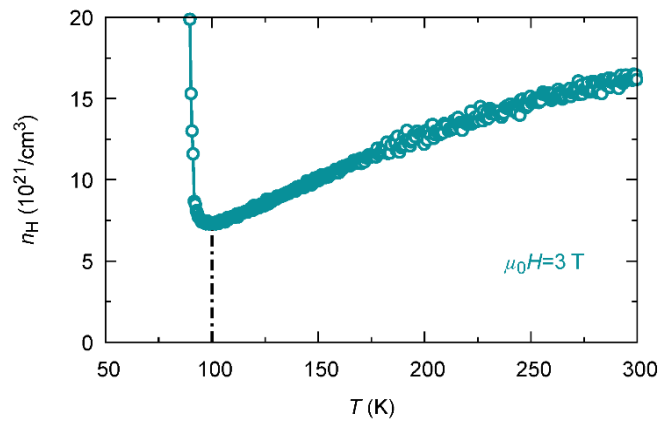

SI-Figure 2: **Temperature dependent charge carrier density:** From Hall effect measurements at constant field (e.g. $\mu_0 H = 3\,T$), the single band charge carrier density of YBCO was obtained via $n_\mathrm{H}(T) = \frac{1}{eR_\mathrm{H}(T)}$ as a function of temperature. Typically in YBCO single crystals, an anisotropy factor is introduced in the determination of the charge carrier density via the Hall constant, accounting for the contribution of the CuO-chains short-circuiting the Hall voltage along the b-direction. However, as here studied films are highly twinned and we do not observe in-plane anisotropy in our Hall measurements, we can safely neglect the effect of the metallic CuO-chains in the calculation of the charge carrier density.

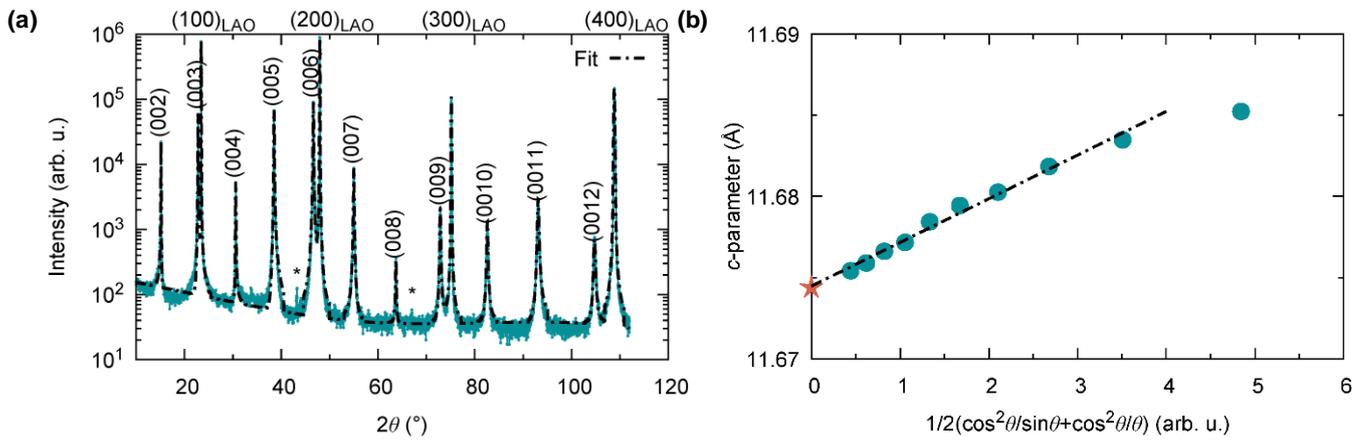

*SI-Figure 3: **c-parameter evaluation via HR-XRD:** (a) X-ray diffraction pattern of YBCO thin film (200 nm) on top of LAO (100) single crystal substrate. Substrate and film peaks can be very well reproduced using Voigt profiles. The background is substracted using a third order polynomial over the full range. Higher order reflection of the substrate peaks (h00) are marked with *. (b) Film c-parameter is obtained by extrapolation of $c = \lambda/2 \sin\theta$ to the intersect $2\theta = 180°$ via a linear fit using the Nelson – Riley formula ($c_{Film}$ = 11.674 Å, as indicated with the red star). This method allows the determination of the lattice parameter with very high precision, since aberration errors are minimised at very high angles.*

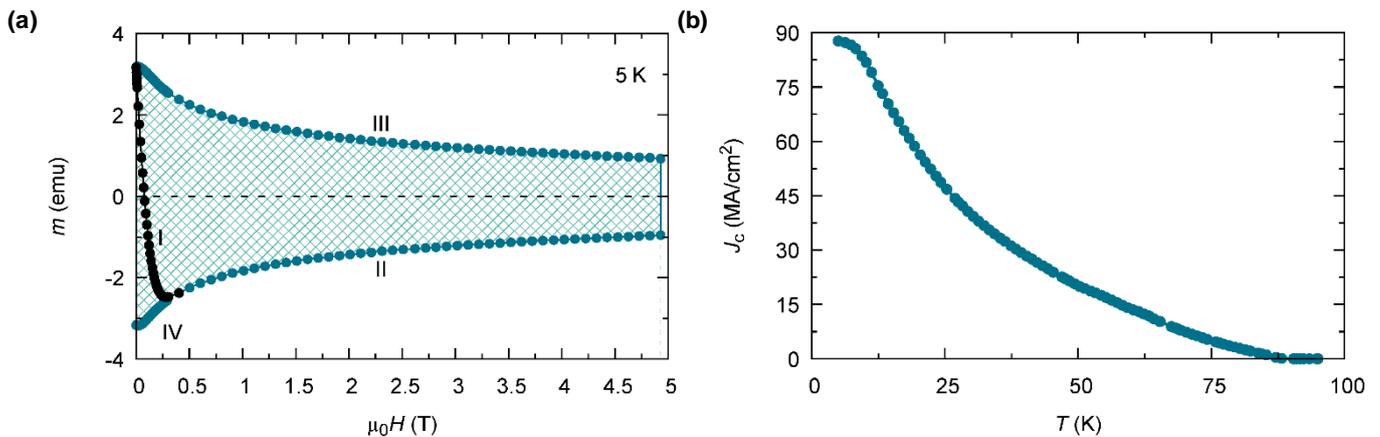

*SI-Figure 4: **Critical current density by SQUID remanent magnetisation measurements:** (a) Irreversible magnetic moment in the saturated $m - H$ hysteresis loop (blue dots) at 5 K. The area between the two branches, corresponding to increasing and decreasing magnetic field, is proportional to the critical current density, $J_c(H)$, through the Bean critical state model. The $m(H)$ hysteresis loop is determined starting from the initial magnetic moment branch (I), by increasing the field up to 5 T (II), subsequent decreasing to -5 T (III), followed by an increase (IV) to 0.3 mT to close the fully saturated loop. (b) Critical current as a function of temperature as obtained by SQUID measurements using Bean critical state model for a thin disc.*

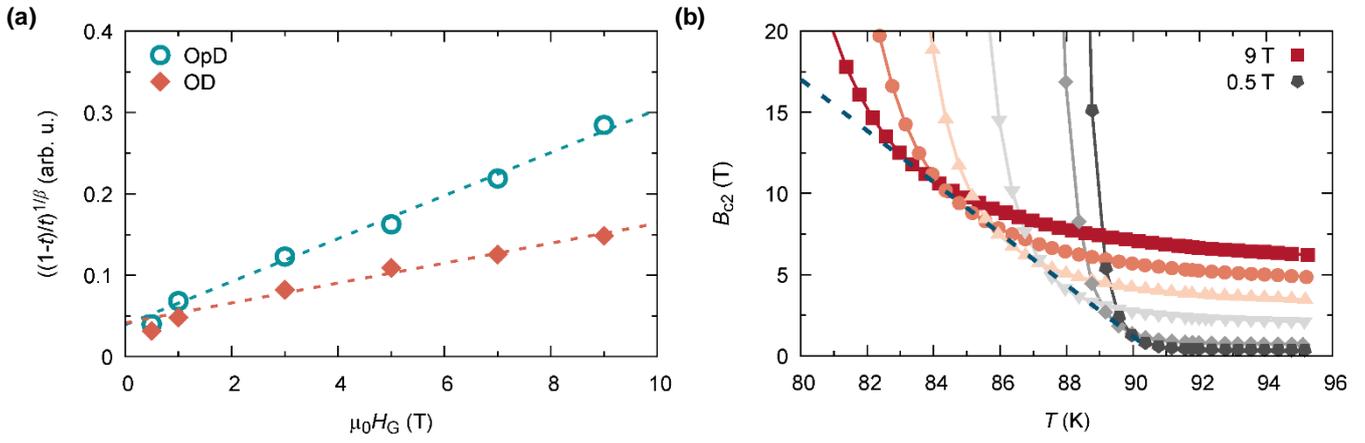

SI-Figure 5: **Analysis of the pinning energy by electrical measurements:** (a) Vortex glass transition line $H_G = H_0 \left[\frac{1-t(H)}{t(H)}\right]^{1/\beta}$ used to extract characteristic magnetic field, $H_0$, via linear fitting, as shown for an optimally and highly overdoped YBCO thin film, giving $\mu_0 H_0 = 37\ T$ and $\mu_0 H_0 = 75\ T$, respectively. $t = T_0(H_G)/T_c$ is the reduced zero-resistance temperature obtained by $\rho(H,T)$ measurements in Van-der-Pauw configuration. $\beta = 0.98$ is a model parameter as reported in literature. (b) The upper critical field close to $T_c$ is determined by the envelope of $\rho_{ff}(H,T)/\rho_N(T) = H/c_{ff}H_{c2}(T)$ curves (dashed line), from which we have obtained the zero temperature limit using the classical Werthamer-Helfand-Hohenberg relation $H_{c2}(0) = -0.69\, T_c \frac{\partial H_{c2}}{\partial T}|_{T_c}$. The coherence length, $\xi(0)$, is then given by $\mu_0 H_{c2}(0) = \phi_0/2\pi\xi(0)^2$. The linear normal-state resistivity, $\rho_N(T)$, is determined above 100 K in zero field and extrapolated to low temperatures, $\rho_{ff}(H,T)$ is the flux flow resistivity and $c_{ff} = 1.45$ is a constant.